\documentstyle[12pt,epsf]{article}

\def \dfrac #1#2 {\displaystyle\frac{#1}{#2}}
\def\be{\begin{eqnarray}}
\def\ee{\end{eqnarray}}
\def\bq{\begin{equation}}
\def\eq{\end{equation}}
\def\ben{\begin{enumerate}}\def\een{\end{enumerate}}

\def\prl {Phys. Rev. Lett. }\def\pr{Phys. Rev. }
\def\np{Nucl. Phys. }\def\pl{Phys. Lett. }

\def\roughly#1{\mathrel{\raise.3ex\hbox{$#1$\kern-.75em%
\lower1ex\hbox{$\sim$}}}}

\begin{document}
\begin{titlepage}
\hfill \today

\vspace{.3cm}
\hfill {\large FTUV 97-29; IFIC 97-29}
\vspace{.2cm}
\begin{center}
\ \\
{\LARGE \bf A quark model analysis 
\\
\vspace{.1cm}
of the transversity distribution$\dagger$}
\ \\
\ \\
\vspace{1.0cm}
{Sergio Scopetta and Vicente Vento$^{(a)}$}
\vskip 0.5cm
{\it Departament de Fisica Te\`orica}

{\it Universitat de Val\`encia}

{\it 46100 Burjassot (Val\`encia), Spain}
 
            and

{\it (a) Institut de F\'{\i}sica Corpuscular}

{\it Consejo Superior de Investigaciones Cient\'{\i}ficas}
\vskip 1cm
\leftline{Pacs: 12.39-x, 13.60.Hb, 13.88+e}
\leftline{Keywords:  hadrons, partons, transversity, evolution.} 
\end{center}
\vskip 1.0cm
\centerline{\bf Abstract}
\vskip 0.4cm
The feasibility of measuring chiral-odd parton distribution functions in
polarized Drell-Yan and semi-inclusive experiments has renewed theoretical 
interest in their study. Models of hadron structure have proven
succesful in describing the gross features of the chiral-even structure
functions. Similar expectations support our study of the
transversity parton distributions in the Isgur-Karl and MIT bag models.
We confirm the diverse low x behavior of the transversity and spin structure 
functions at the experimental scale and show that it is fundamentally a 
consequence of the different behavior under evolution of these functions. 
The  inequalities of Soffer establish constraints between data and model 
calculations of the chiral-odd transversity function. The approximate 
compatibility of our model calculations with these constraints 
conferes credibility to our estimates.

\vspace{.7cm}

{\tt
\leftline {scopetta@titan.ific.uv.es}
\leftline{vicente.vento@uv.es}}
\vspace{0.7cm}
\noindent{\small$\dagger$Supported in part by DGICYT-PB94-0080, DGICYT-PB95-0134 
and TMR programme of the European Commisison ERB FMRX-CT96-008}
\end{titlepage}

\section{Introduction}
\indent\indent
The parton distributions on the lightcone are physical quantities which 
describe the low-energy properties of the nucleon in high-energy processes.
One may define for a hadron an infinite number of parton distributions, however
in high-energy processes we are interested in those with low twist. At the
twist two level the quark parton model defines three parton
distributions labelled conventionally as $f_1(x)$, $g_1(x)$ and
$h_1(x)$. The first two have been the main focus of studies so far. 
The last has received little attention because it plays a minor role in
deep inelastic processes. Not until the work of Ralston and Soper
\cite{rs79} its importance in characterizing the nucleon´s high-energy 
properties was recognized. Jaffe and Ji \cite{jj91} named it transversity 
distribution because it measures the density of quarks (minus antiquarks) in 
eigenstates of the transverse Pauli-Lubanski operator. 

$h_1$ is a twist two parton distribution which, unlike the others, is
chiral-odd. This makes it unaccessible to conventional inclusive DIS 
experiments except if the current quark masses are considered, in which case
arises as a twist three contribution \cite{ko79}. However it can be
measured at the leading twist in the Drell-Yan process with both beam
and target polarized \cite{rs79,cp91}. Recently, semi-inclusive lepton
nucleon DIS processes have been proposed as a way of determining $h_1$
\cite{am90,co93,jj94,tm95}. The time has come to investigate the chiral-odd
and higher twist parton distribution functions in detail since  measurements
will appear in the next years from RHIC \cite{yo95}, HERA \cite{an96}, 
CERN \cite{bp96} and from new facilities such as ELFE \cite{la97}.
The possibility of measuring $h_1$ has renewed theoretical interest and
estimates to guide the experimental analysis have been presented using
various methods, i.e., leading log approximation of $QCD$ \cite{km96},
$QCD$ sum rule approach \cite{ik94,hj95} and model calculations
\cite{jj91,st93,bc96}.  

The aim of this paper is to study the transversity distribution within 
different models of hadron structure. As pointed out by Jaffe and Ross 
\cite{jr80}, these calculations are associated with a low $Q^2$, 
the so called hadronic scale. To go from the hadronic scale to the 
experimental conditions our scheme proceeds via perturbative $QCD$ evolution. 
It has been claimed in the past that $h_1 \approx g_1$, a result which
arises naturally in model calculations. We will show that the very diverse 
evolution properties of these two structure functions lead, even if they are
similar at the hadronic scale, to large differences at the usual 
experimental conditions, a result known to other authors \cite{st93,bc96}. 
This confirmation within different schemes leads to an optimal experimental 
scenario.

Soffer \cite{so95} has produced a 
series of inequalities in the parton model, relating the transversity 
distributions with the chiral-even distributions. When combined with
data, these inequalities provide rigorous
constraints for model calculations of $h_1$. We analyze their applicability 
and  obtain the limitations of the models under scrutiny.

\section{The theoretical framework}
The transversity structure function measures the polarization asymmetry of
quarks (or antiquarks) in a transversly polarized hadron, i.e.,
\bq
h_1(x,Q^2) =\frac{1}{2} \sum_q e^2_q (h^q_1(x,Q^2) + h^{\bar{q}}_1(x,Q^2)),
\eq
where the transversity parton distribution functions are given by
\bq
h^q_1(x,Q^2) = q_{\uparrow}(x,Q^2) -q_{\downarrow}(x,Q^2).
\eq
Here $\uparrow (\downarrow)$ indicates that the spin of the quark of 
flavor $q$ is parallel (antiparallel) to the tranverse polarization of
the nucleon and $\bar{q}$ refers to the equivalent antiquark distributions.
   
    The following inequalities arise  from current-hadron 
amplitudes,
\bq
|g_1(x,Q^2)| \leq f_1(x,Q^2)
\eq  
and
\bq
|h_1(x,Q^2)| \leq f_1(x,Q^2),
\eq
which  also appear as a trivial consequence of the definition of
the distribution functions in the lightcone 
helicity and transversity bases, respectively \cite{jj91}. 

A third inequality was proven by Soffer for the parton model \cite{so95,gj95},
\bq
2|h_1^q(x,Q^2)| \leq g_1^q(x,Q^2) + f_1^q(x,Q^2)
\label{soffer}
\eq
and its behavior under evolution has been clarified by Barone \cite{ba97}.

The structure function $h_1(x,Q^2)$ is associated with the tensor operator 
$\bar{\Psi}\sigma_{\mu\nu}i\gamma_5\Psi$. The $n=1$ sum rule leads to
\bq
\delta q(Q^2) =\int^1_0 dx((h^q_1(x,Q^2) - h^{\bar{q}}_1(x,Q^2))
\label{tcharge1}
\eq
where $\delta q$ is named the {\it tensor charge} and is given in the
nucleons's rest frame by
\bq 
<PS|\bar{\Psi}_q\Sigma_i \Psi_q|PS> = 2\delta q S_i
\label{tcharge2}
\eq
where $\Psi_q$ here labels the spinor of flavor q, $\Sigma_i$ is the 
conventional spin operator and $S_i$ the nucleon's spin
vector defined as usual \cite{jj91}. From eq.(\ref{tcharge1}) we see that
the tensor charge counts the number of valence quarks of opposite
transversity. Since $\delta q $ is charge conjugation-odd, it gets no
contribution from quark-antiquark pairs of the sea.

In contrast the quark spin operator associated with $g_1$ is even under
charge conjugation and therefore the corresponding equation is
\bq
\Delta q(Q^2)=\int^1_0 dx((g^q_1(x,Q^2)) + g^{\bar{q}}_1(x,Q^2)),
\label{scharge1}
\eq
where $\Delta q$, the axial charge, is given in the nucleon rest frame by
\bq 
<PS|\Psi_q^+\Sigma_i \Psi_q|PS> = 2\Delta q S_i,
\label{scharge2}
\eq
It is evident that $\Delta q$ includes the helicity of the sea. From
eqs.(\ref{tcharge2}) and (\ref{scharge2}) it is apparent that in
non-relativistic model calculations $\delta q =\Delta q$. This argument
can be generalized to all the other moments so that one may prove that
$h_1 = g_1$ in these models.

The evolution properties of $h_1$ and $g_1$ are however very different.
All of the local operators associated with $h_1$ have non-vanishing
leading order anomalous dimensions. Moreover no gluon operators
contribute to $h_1$ in any order because it is chiral-odd and the gluon
operators are all chiral even. Therefore $h_1$ is a non-singlet structure
function, which evolves homogeneously with $Q^2$ and none of its moments
is $Q^2$ independent. On the contrary the non-singlet components of the
first moments of $g_1$ have vanishing anomalous dimensions, thus they are
$Q^2$ independent and their singlet components mix with gluons in a 
complicated way governed by the axial anomaly \cite{am90,co93,co84}. 

The formalism of our calculation was described in detail in ref.\cite{tr97}, 
although we will in some cases diverge from some of its details to respect 
the author's philosophy in the models. We assume that quark model calculations 
give the value of the matrix elements at a definite hadronic scale $\mu_0^2$.
The leading twist contribution of the matrix elements is evolved to
the experimental conditions at high $Q^2$ by means of renormalization group 
methods of perturbative $QCD$. The analysis here will be carried out to
leading order (LO) since the evolution parameters of $h_1$ are only known to
this order \cite{ko79,am90}.

At the hadronic scale the physical meaning of the structure functions is
very intuitive in a naive non-relativistic formulation \cite{tr97}:
\be
g_1^q(x,\mu_0^2) &= & \frac{m_q}{M} \int d^3p (n_q^{\rightarrow} - 
n_q^{\leftarrow}) \delta\left(x-\frac{p^+}{M}\right)\\
h_1^q(x,\mu_0^2) & = & \frac{m_q}{M} \int d^3p (n_q^{\uparrow} - 
n_q^{\downarrow}) \delta\left(x-\frac{p^+}{M}\right)\\
\ee
where the spin dependent momentum distributions of the quark $q$ are
\be
n_q^{\rightarrow (\leftarrow)}(\vec{p}) & = &<P S_z|\sum_{i=1}^3 P^q_i 
\frac{1+(-)\sigma_i^z}{2}|P  S_z>\\
n_q^{\uparrow (\downarrow)}(\vec{p})& = &<P S_x|\sum_{i=1}^3 P^q_i 
\frac{1+(-)\sigma_i^x}{2}|P  S_x>\\
\ee
Here $P^q_i$ is the flavor projection operator. It is clear that due to
rotational invariance the two matrix elements are identical, i.e.,
$h^q_1(x,\mu_0^2) = g^q_1(x,\mu_0^2)$. This is not the case in
relativistic models where the contribution from the lower components
makes them different. In the latter case we will use \cite{jj91,st93,ja75}
\be
g_1(x,\mu_0^2) & = & \int\frac{d\lambda}{4\pi}
e^{i\lambda x} <P S_z|\bar{\Psi}(0)\;\backslash\hspace{-2.5mm}{n}
\gamma_5 \Psi (\lambda n)|_{\mu^2_0}|P S_z>\\
h_1(x,\mu_0^2) & = & \int\frac{d\lambda}{8\pi}
e^{i\lambda x} <P S_{\perp}|\bar{\Psi} (0)[\;\backslash\hspace{-3mm}
{S_{\perp}},\; \backslash\hspace{-2.5mm}{n}]\gamma_5 
\Psi(\lambda n)|_{\mu^2_0}|P S_{\perp}>
\ee
where $n^{\mu}=\frac{1}{\sqrt{2} p} (1,0,0,-1)$.
One should realize that these equations also incorporate the relativistic
corrections to the non-relativistic calculations.

These set of equations represent the basis for the calculation of matrix 
elements at the hadronic scale. The sheme is completed by evolving using
the conventional procedure developed for this problem in ref.\cite{tr97}.

\section{Results}
\indent\indent
We will discuss the results in two models
\begin{itemize}
\item[i)] The non relativistic model of Isgur-Karl \cite{ik78};
\item[ii)] The relativistic MIT bag model \cite{ja74}.
\end{itemize}

In both cases we will use the corresponding support correction as defined in
\cite{tr97} and \cite{jr80}, respectively. In Figs. 1 we show the results 
corresponding to $g_1$ and $h_1$ for the first case. Fig. 1a corresponds
to the pure valence quark hadronic scenario of ref.\cite{tr97}, characterized 
by a very low hadronic scale ($\mu_0^2 = 0.079 GeV^2$). Fig. 1b corresponds to 
the second scenario of ref.\cite{tr97}, where $40 \% $ of the momentum is carried by valence
gluons and therefore the hadronic scale is  larger ($\mu_0^2 = 0.3
GeV^2$). The initial data are evolved to $10 GeV^2$. 

\begin{figure}[htb]
\vspace{6cm}
\includegraphics{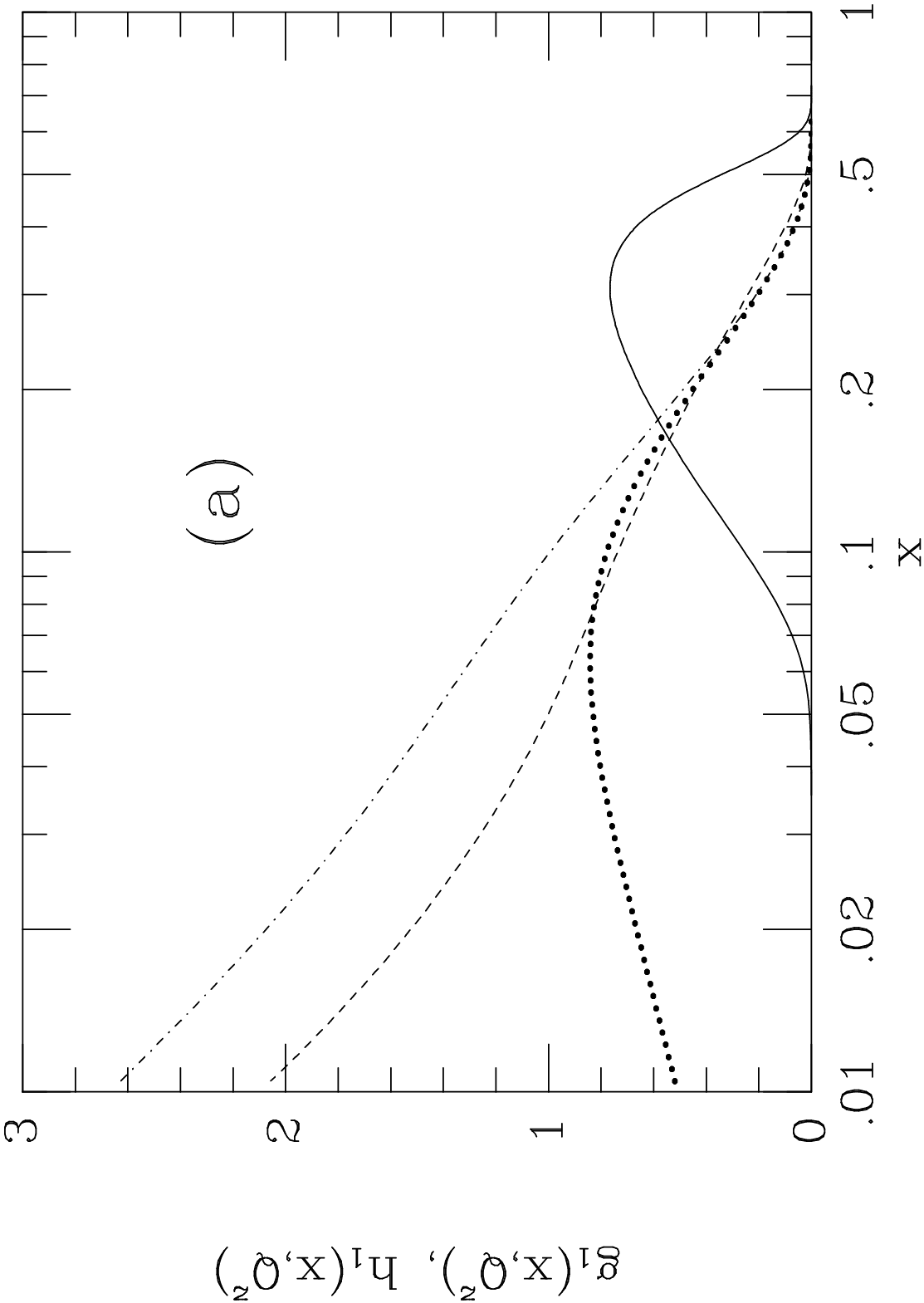}
\includegraphics{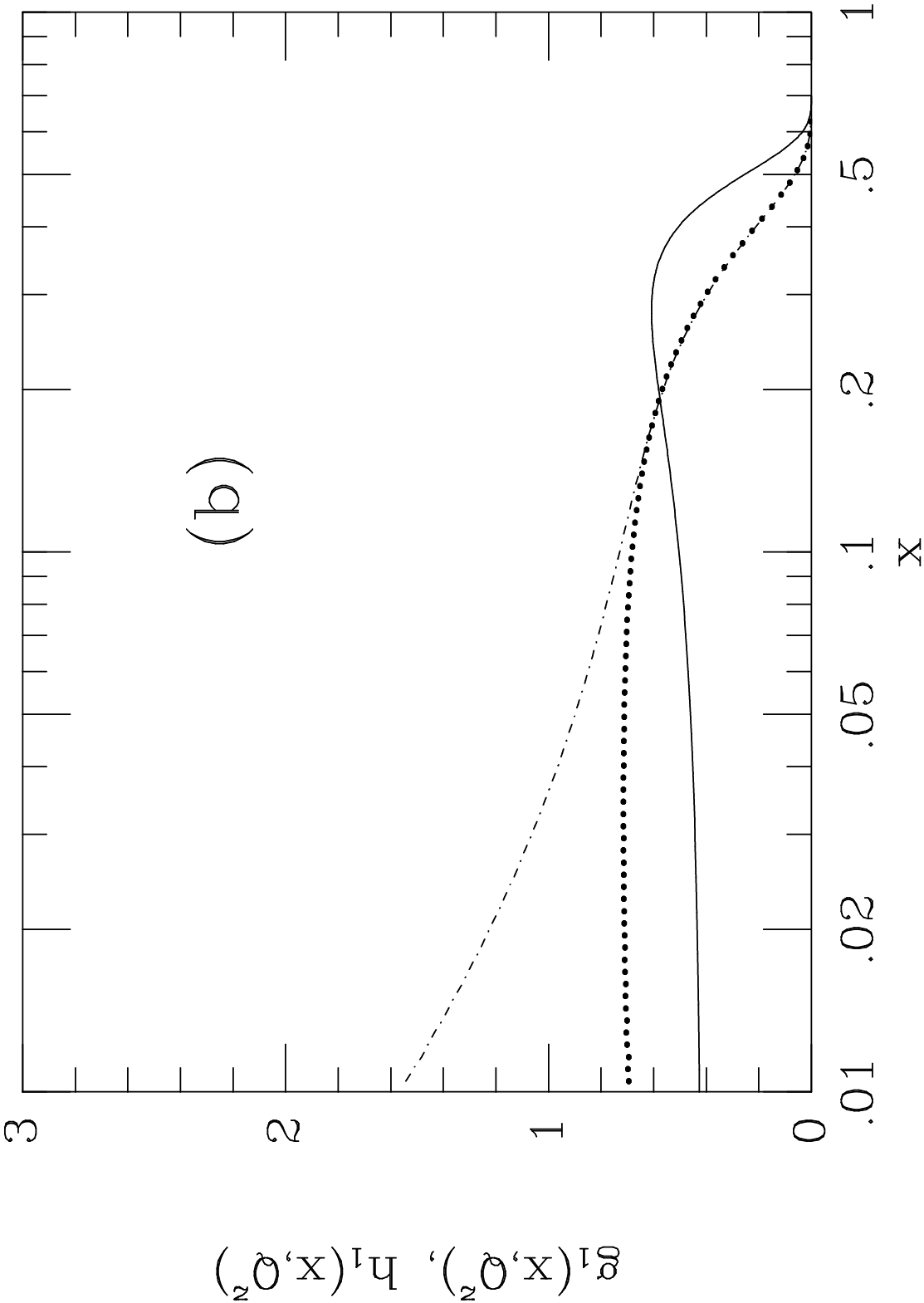}
{\small {\bf Figure 1}: We show the trasversity function $h_1(x,\mu_0^2)$ 
(continuous line) which coincides with the spin distribution function 
$ g_1(x,\mu_0^2)$  for the a) Isgur-Karl model \cite{ik78} at the hadronic 
scale $\mu_0^2 =0.079 GeV^2$; b) Isgur-Karl based model with $40\%$ valence 
gluons at the hadronic scale $\mu_0^2 =0.3 GeV^2$ \cite{tr97}.
Their corresponding evolved (LO)  distributions $h_1(x,Q^2)$ (dotted) and 
$g_1(x,Q^2)$ (dot-dashed) at $Q^2= 10GeV^2$ are also shown. In Fig a) we  
draw the (NLO) evolution of the spin structure function for comparison 
(dashed line).}
\end{figure}

The figures show that the diverse evolution properties of these two structure 
functions lead to a large difference between the two initially identical 
functions. The difference occurs at small $x$ and has been noted also by other 
authors \cite{bc96,ba97}. In Fig. 1a we  show the Next to Leading Order 
evolution of $g_1$ as measure of our theoretical uncertainties.   
We do not expect in the case of $h_1$ an erratic NLO 
behavior and  are confident that our discussion prevails beyond 
the leading order approximation.

In Figs. 2 we analyze the result of the same calculation in the support
corrected MIT bag model. It has been argued \cite{ja75} that gluons in the 
bag are not carrying  momentum. This would imply, according to conventional 
$QCD$ wisdom, that the hadronic scale should be very low ($\mu^2_0 \approx 
0.079 GeV^2$)\cite{st93}. On the other hand the comparison of the bag $f_1$ 
moments with the experimental results \cite{jr80} suggests a much larger 
hadronic scale ($\mu^2_0 \approx 0.75 GeV^2$). We show therefore in Figs 2.
both scenarios. 

\begin{figure}[htb]
\vspace{6cm}
\includegraphics{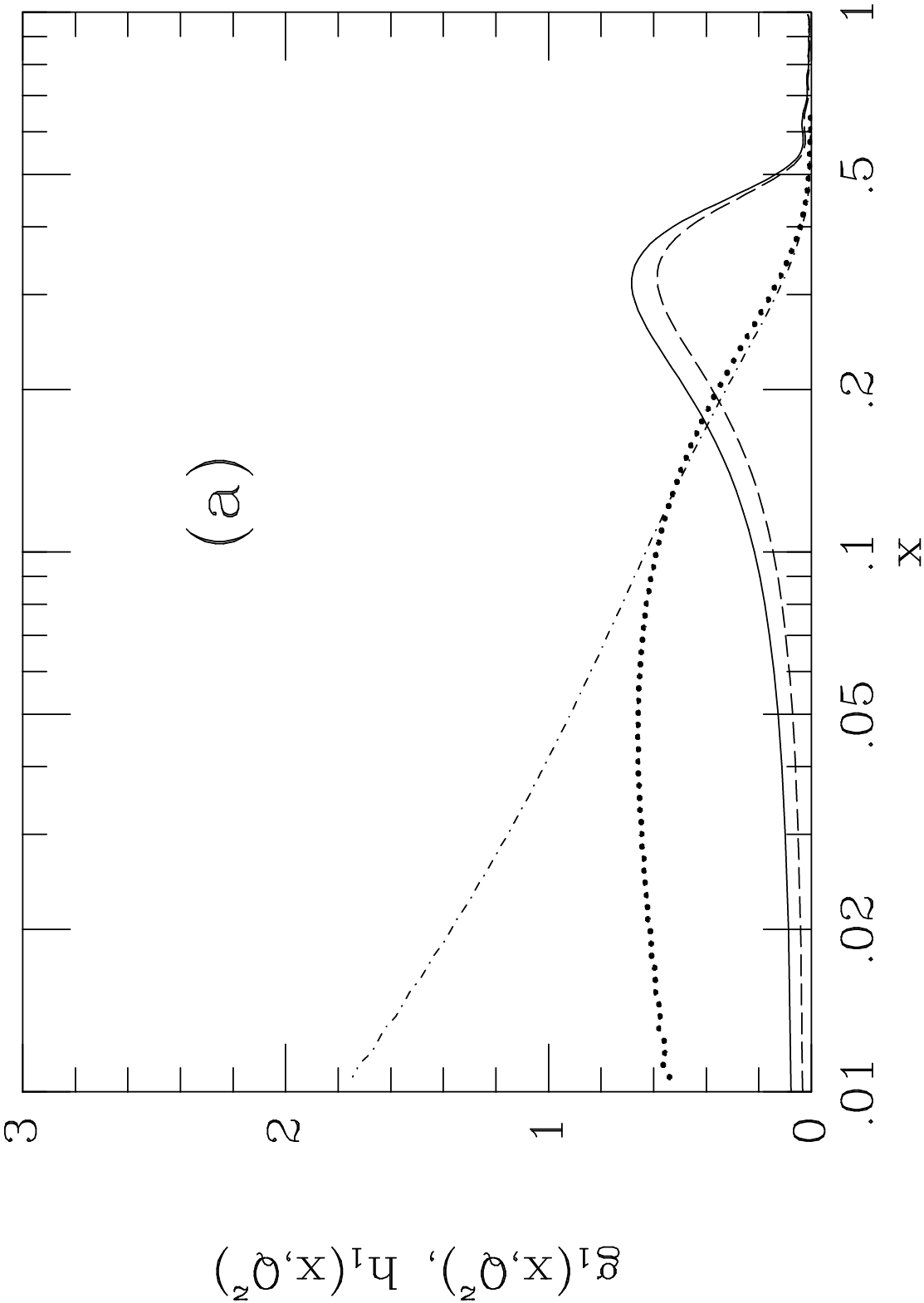}
\includegraphics{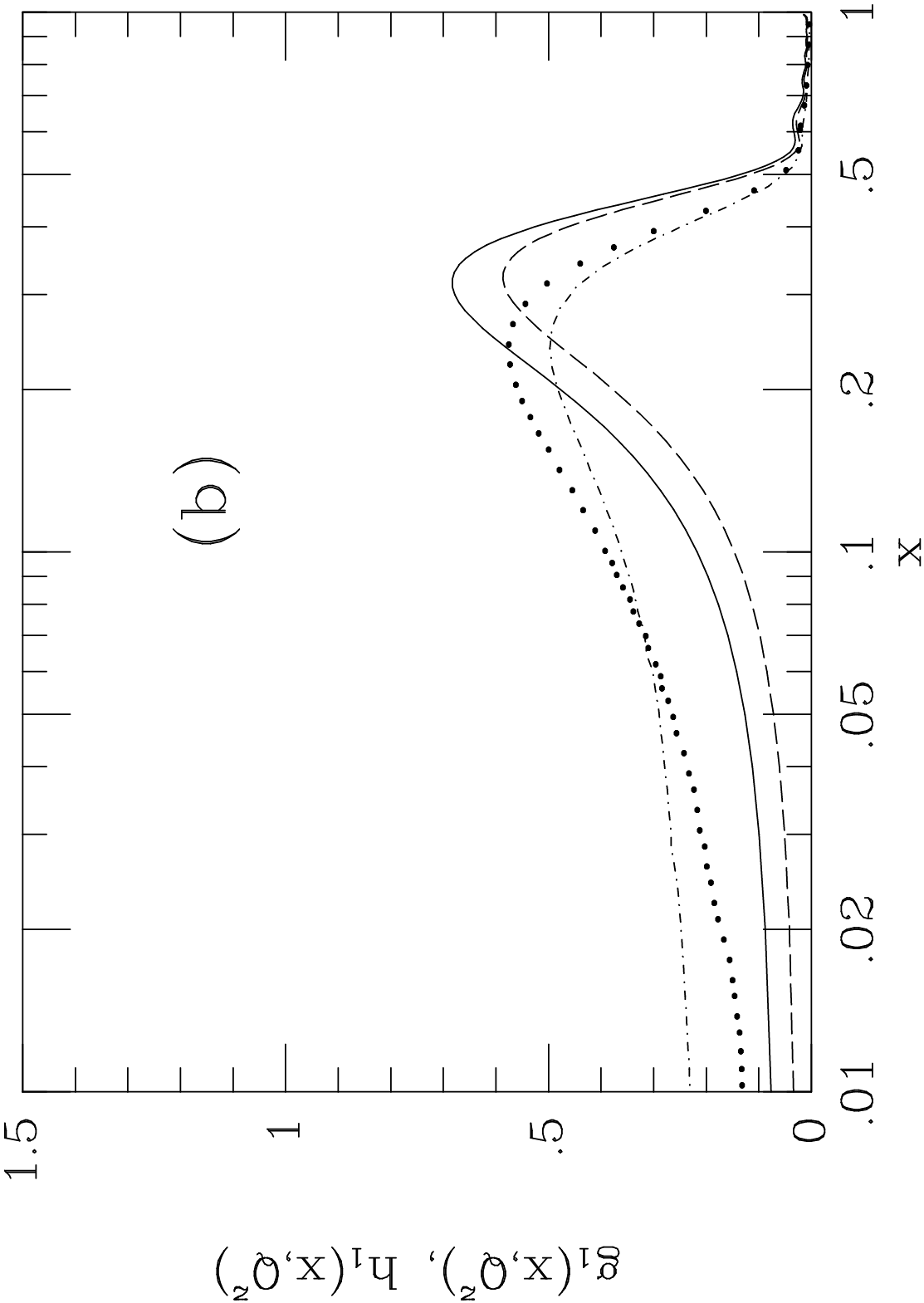}
{\small {\bf Figure 2}: 
The trasversity function $h_1(x,\mu_0^2)$ (continuous line)
and the spin distribution function $g_1(x,\mu_0^2)$ (dashed line) for the 
support corrected MIT bag model \cite{ja74} at the hadronic scale
a) $\mu_0^2 =0.079 GeV^2$ and b) $\mu_0^2 =0.75 GeV^2$ are shown. Their 
corresponding evolved  distributions 
$h_1(x,Q^2)$ (dotted)and $g_1(x,Q^2)$ (dot-dashed) at $Q^2= 10GeV^2$ are also 
shown.}
\end{figure}

This lack of precise definition of the hadronic scale has led to Fig. 3 where
we  analyze the dependence of the difference between $h_1$ and $g_1$ at small 
x as a function of the hadronic scale in the MIT bag model. It is clear from 
the figure that a small hadronic scale implies large differences between the 
two structure functions while the opposite is true in the case of large 
hadronic scales. The hadronic scale controls the magnitude of the evolution
and since the difference arises mostly from evolution, the behavior at small
x of these functions will control this parameter.

\begin{figure}[htb]
\vspace{6cm}
\includegraphics{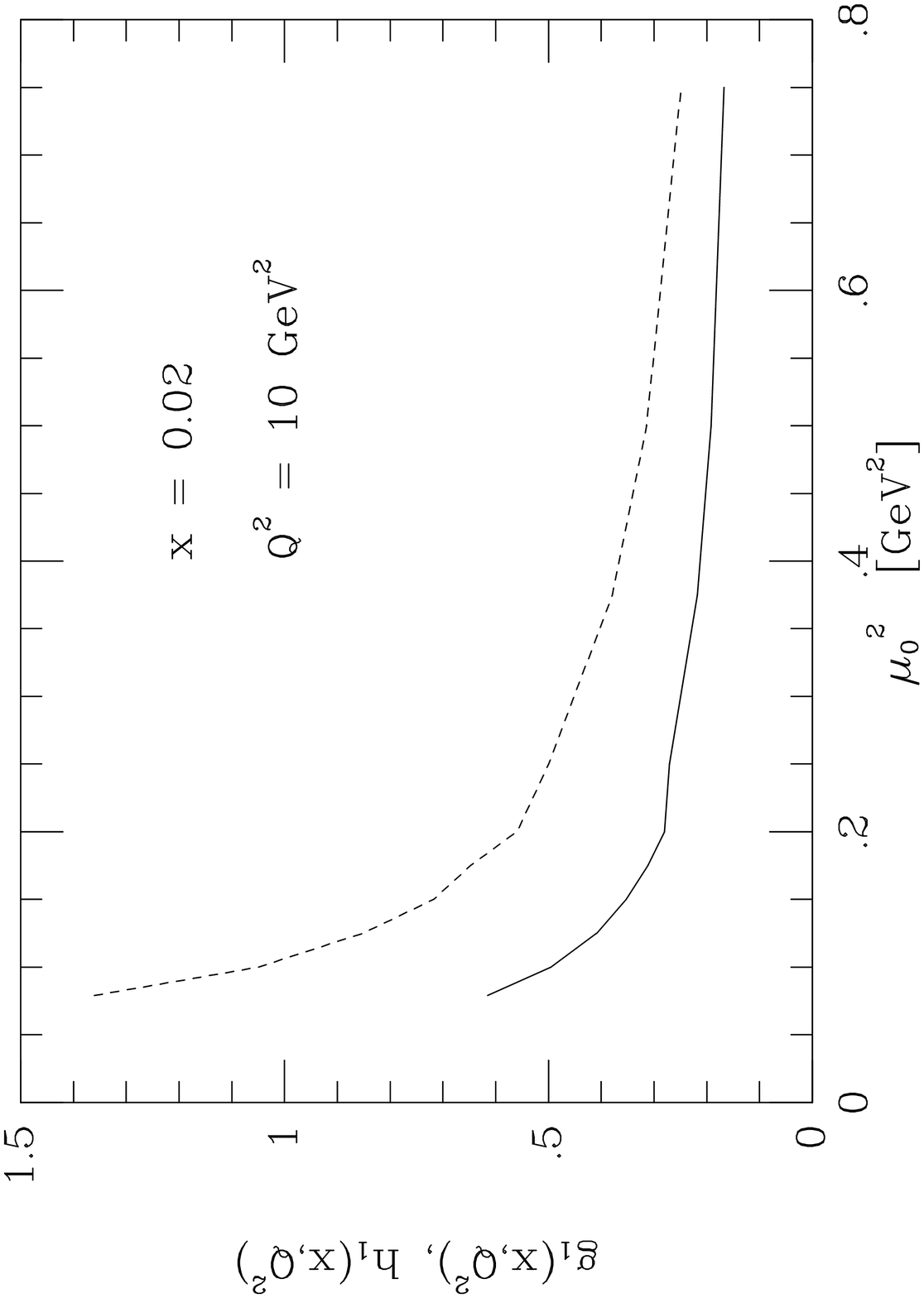}
{\small {\bf Figure 3}: 
We show the value of $h_1$ (full) and $g_1$ 
(dashed) at $x =0.02$ as 
a function of the hadronic scale for the support corrected MIT bag model 
calculation.}
\end{figure}

For completeness 
we show in the table the magnitude of the tensor and axial charges, noting 
that the latter are renormalization group invariant to leading order (
see \cite{hj95} where this calculation is discussed in detail for the MIT
bag model.)

\begin{table}[tbh]
\begin{center}
\begin{tabular}{|c|c|c|c|}\hline
&&$\mu_0^2$&$Q^2$ \\ \hline
$\delta h$& IK & 0.270 &0.183 \\ \cline{2-4} 
& MIT &0.215&0.146  \\ \hline
$\Delta g$& IK& 0.270&0.270  \\ \cline{2-4}
& MIT& 0.176 & 0.176 \\ \hline
\end{tabular}
\end{center}
{\small {\bf Table}: The values of the tensor charge, $\delta h =
\frac{1}{2}\sum_q e_q^2\delta q$, and the spin charge,
$\Delta g=\frac{1}{2}\sum_q e_q^2\Delta q$, are shown, both at the hadronic scale
and at the experimental scale, and for the two models studied. 
L.O. evolution  from a hadronic scale of $\mu_0^2 =0.079 GeV^2$ 
to $Q^2 = 10 GeV^2$ has been performed.}
\end{table} 

To finish this section we turn to Soffer's inequalities.
Both models verify the primitive Soffer inequality eq.(\ref{soffer}), not only
at the hadronic scale, but also as we evolve the distribution function towards
the physical regime. However Soffer argued \cite{so95} that his positivity 
bound could be used combined with data to limit the validity of models. In
particular by imposing the simple relation 
\bq
\Delta u(x) = u(x) - d(x)
\eq
proposed in \cite{bs95} and which is well supported by the data, it is possible
to use the positivity bound  to obtained the allowed range of values
for $h^u_1$, namely
\bq
u(x) - d(x)\ge |h^u_1(x)|
\label{soffer1}
\eq
The MIT bag model  fails the bound for large values of x \cite{so95}.

\begin{figure}[htb]
\vspace{6cm}
\includegraphics{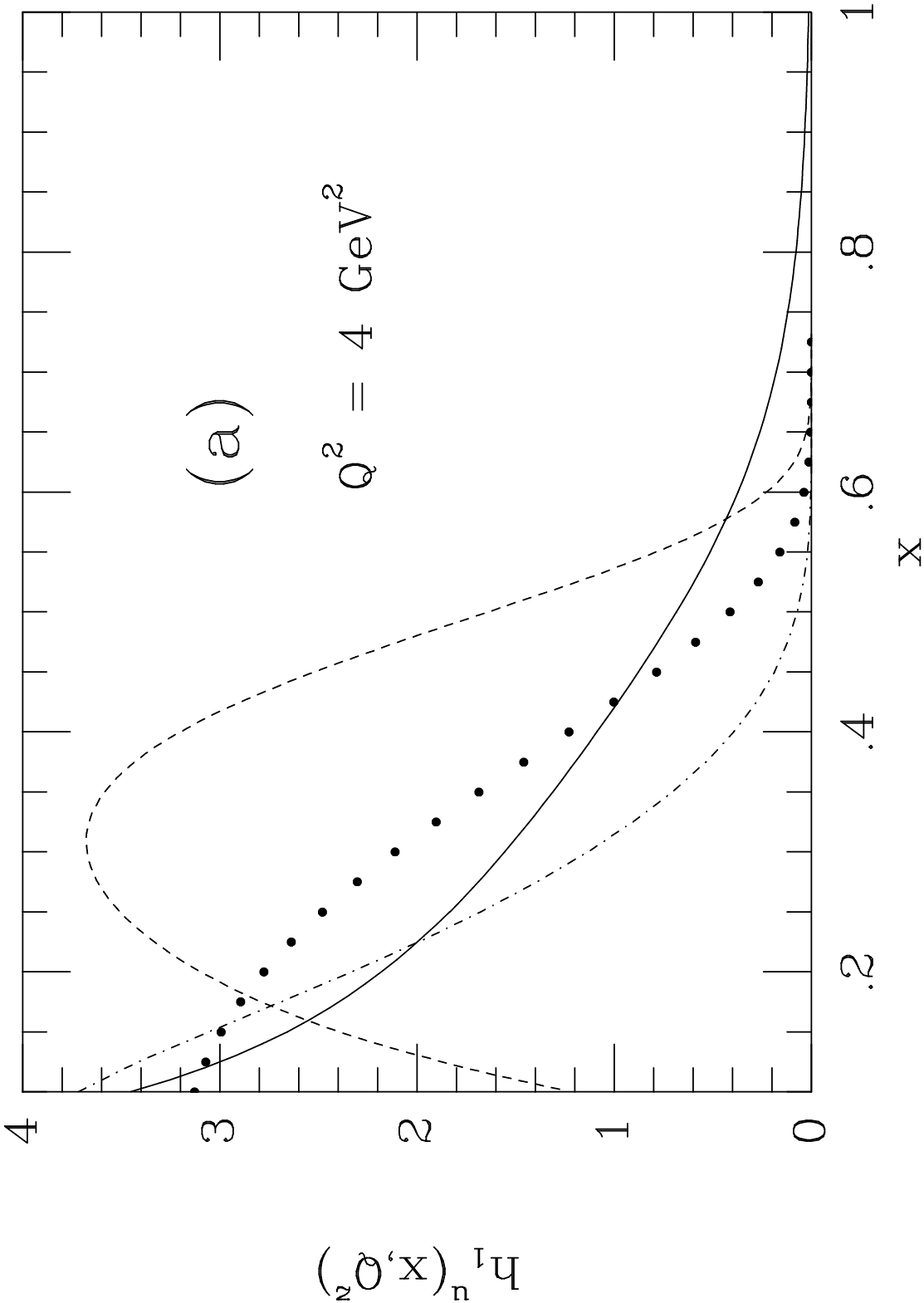}
\includegraphics{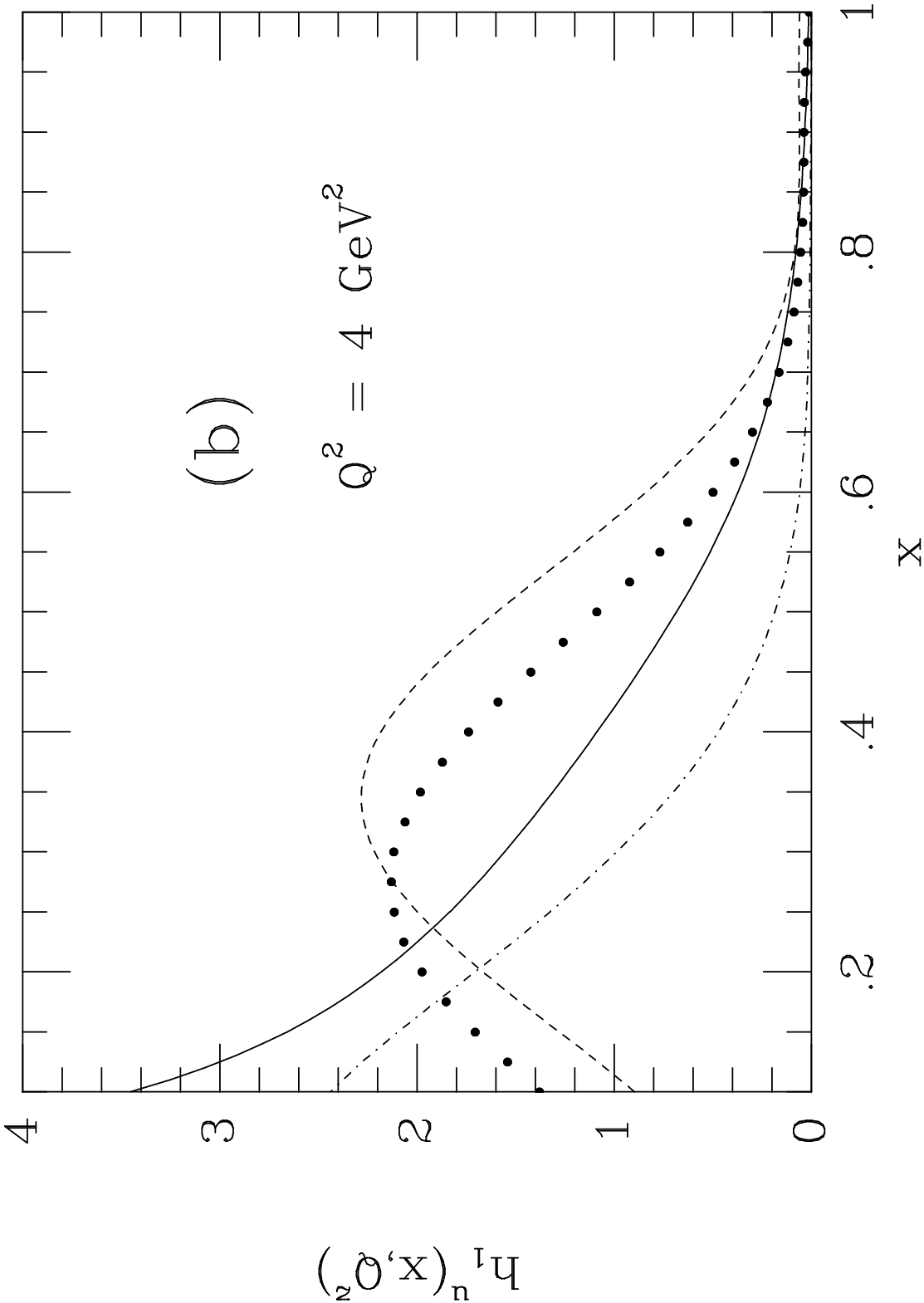}
{\small {\bf Figure 4}: 
The allowed region of Soffer determined from the experimental 
data at $4 GeV^2$, corresponds to the region inside the continuous line.  
Fig.a): the dashed line corresponds to the pure Isgur-Karl model calculation 
\cite{ik78};the dot-dashed line represents the evolved IK solution from an
hadronic scale of $0.079 GeV^2$, while the dotted line corresponds to the evolved
solution of the IK model supplemented by gluons from an hadronic scale of 
$0.3 GeV^2$. Fig b): the dashed line corresponds to 
the pure MIT calculation; the dotted line assumes a hadronic scale of $0.75 
GeV^2$, while the dot-dashed line one of $0.079 Gev^2$, consistent with the 
Stratmann analysis \cite{st93}. The evolution has been carried out to leading 
order.}
\end{figure}

We show in Fig. 4 the comparison of the experimental constrain at $4 GeV^2$ 
with the Isgur-Karl and MIT bag calculations. In the figure the allowed region 
is described by taking the lefthand side of eq.(\ref{soffer1}) from the data. 
The remaining curves represent the righthand side of the equation which 
we have calculated from the models. It is clear from the figure that at the  
hadronic scale, neither fulfils the constraint. Since these inequalities should
be  valid in the partonic regime \cite{gj95}, we show the values resulting from
evolving the model calculations from the hadronic scale to the scale of the 
data ($4 GeV^2$) As the figure shows, greater consistency is achieved 
after this procedure. Moreover the analysis of Bourrely ans Soffer \cite{bs95}
pays no attention to the possible errors associated with the experimental fit.
If these are taken into account the consistency is even better. 
This result does not imply that the conventional models of hadron structure
taken as a description of the physics at the hadronic scale are quantitatively 
succesful in explaining the deep inelastic data. As stated in previous analysis
\cite{tr97}, these models give a qualitative description, which we have confirmed
for the Soffer inequalities. However in order to obtain a quantitative 
description additional ingredients have to be added. By looking at  Fig. 4 we
rediscover the need for high momentum components in the Isgur-Karl model.
Moreover the inclusion of gluons allows compatibility with a much larger
hadronic scale. The same figure teaches us that the Stratmann scenario 
\cite{st93} with small hadronic scale for the MIT bag model is better realized
then the large hadronic scale scenario of Jaffe and Ross \cite{jr80}.

\section{Conclusions}
\indent\indent

The use of models of hadron structure to describe the deep inelastic properties
of the proton and neutron has proven successful for the chiral-even twist two 
structure functions \cite{tr97} (and references therein). Several authors have 
generalized  the analysis  to the transversity functions \cite{jj94,st93,bc96}. 
Since these have not been measured, this analysis has the added value of 
prediction. We have completed the spectrum of possible calculations by including 
that of a well established not relativistic model, with a fine tuned technique 
for constructing the structure functions and performing the RGE evolution. 
Moreover we have returned to the highly succesful field theoretic approach of
the MIT bag model and reanalized some of the features questioning its validity, 
i.e., the highly discussed Soffer inequalities.

The features of the analysis which we next discuss are herewith well established.
The evolution properties of the $h_1$ and $g_1$ structure functions is the main
ingredient which distinguishes between them. We have analyzed the evolution from
the hadronic scale to the experimental scale. In this scenario, even if the two 
structure functions are similar at the hadronic scale, the evolution makes them 
very different at the experimental scale. Moreover their difference occurs at 
small x. We can see that the quantitative difference between the various models 
models is not large, therefore we are confident in the size of the estimates. 

We have studied the dependence on the hadronic scale. This parameter turns out
to be a main ingredient of the description since it controls the magnitude of
the evolution process. In the case of the MIT bag model a small hadronic scale,
as used for example by Stratmann \cite{st93} leads to considerable differences
between the two structure functions at small x, while the large one proposed 
by Jaffe and Ross \cite{jr80} implies small, but still detectable differences.

The above analysis leads to an experimental scenario characterized by precise
measurements in the small x region. The magnitude of the structure functions 
in that region is a measurement of the hadronic scale, while that of their 
difference gives more indication about the internal structure of the hadron.

Once it is realized that Soffer´s inequalities are statements valid at the 
partonic scale and not at the hadronic scale, even if, as shown by Barone 
\cite{ba97}, they are respected by evolution, the models under scrutiny in
this paper do satisfy them approximately as long as the hadronic scale is taken
in the strict sense \cite{st93,tr97}.

\section*{Acknowledgements}
\indent\indent
We thank X. Artru for useful correspondence
regarding the evolution of the transversity structure function.  
We have mantained iluminating discussions with Marco Traini and acknowledge
some tuition on the evolution code.

\end{document}